\documentclass[conference]{IEEEtran}

\usepackage[english]{babel}
\usepackage{blindtext}
\usepackage{xcolor}
\usepackage[utf8]{inputenc}

\usepackage[nomain, toc, acronym]{glossaries}
\usepackage{subfig}
\usepackage{hyperref}
\usepackage{amsmath}
\usepackage{amssymb}
\usepackage{graphicx}
\usepackage{booktabs}
\usepackage{multirow}
\usepackage{float}

\newacronym{dl}{DL}{Deep Learning}
\newacronym{rl}{RL}{Reinforcement Learning}
\newacronym{ours}{LFQ}{Learning Fair Qdisc}
\newacronym{cca}{CCA}{Congestion Control Algorithm}
\newacronym{aqm}{AQM}{Active Queue Management}
\newacronym{rtt}{RTT}{Round Trip Time}
\newacronym{fq}{FQ}{Fair Queuing}
\newacronym{ml}{ML}{Machine Learning}
\newacronym{bdp}{BDP}{Bandwidth Delay Product}
\newacronym{mae}{MAE}{Mean Absolute Error}
\newacronym{mse}{MSE}{Mean Squared Error}
\newacronym{qoe}{QoE}{Quality of Experience}

\begin{document}
\title{LFQ: Online Learning of Per-flow Queuing Policies using Deep Reinforcement Learning}

\author{\IEEEauthorblockN{Maximilian Bachl, Joachim Fabini, Tanja Zseby}
\IEEEauthorblockA{Technische Universität Wien\\
Vienna, Austria\\
firstname.lastname@tuwien.ac.at}}

\maketitle

\begin{abstract}
The increasing number of different, incompatible congestion control algorithms has led to an increased deployment of fair queuing. Fair queuing isolates each network flow and can thus guarantee fairness for each flow even if the flows' congestion controls are not inherently fair. So far, each queue in the fair queuing system either has a fixed, static maximum size or is managed by an Active Queue Management (AQM) algorithm like CoDel. In this paper we design an AQM mechanism (Learning Fair Qdisc (LFQ)) that dynamically learns the optimal buffer size for each flow according to a specified reward function online. We show that our Deep Learning based algorithm can dynamically assign the optimal queue size to each flow depending on its congestion control, delay and bandwidth. Comparing to competing fair AQM schedulers, it provides significantly smaller queues while achieving the same or higher throughput. 
\end{abstract}

\section{Introduction}
\label{sec:intro}

New \glspl{cca} for TCP and QUIC are still being introduced, revisiting old ideas but also introducing new concepts \cite{dong_pcc_2018, cardwell_bbr:_2016, hock_tcp_2017, bachl_rax_2019, jay_deep_2019}. Some of them, such as BBR, are already being used in real-world production systems. While new \glspl{cca} are commonly being designed with compatibility to other \glspl{cca} in mind, often they do not share the link completely fairly with older \glspl{cca} \cite{hock_experimental_2017, fejes_who_2019, fejes_incompatibility_2020}. Besides that, network flows using the same \gls{cca} can also be unfair to each other: For example, BBR favors flows with a high \gls{rtt}, while New Reno favors those with a low one \cite{turkovic_interactions_2019,turkovic_fifty_2019}. This unfairness can be mitigated by using \gls{fq} at the bottleneck link, isolating each flow from all other flows and assigning each flow an equal share of bandwidth \cite{dumazet_pkt_sched:_2013}. 

Given the high throughput and low queuing delay that is required by emerging cloud gaming \cite{jarschel_evaluation_2011} and virtual reality \cite{elbamby_toward_2018} applications, the optimal management of each flow's queue becomes an increasingly relevant question. Furthermore, for 5G, ultra low latency communications have been made a priority \cite{li_5g_2018} and we argue that effective buffering must also be taken into account for achieving this goal. The simplest solution manage a flow's buffer is to use a static buffer size (queue size) for each flow. However, this can lead to inacceptably large queuing delays. Another solution is to use an advanced \gls{aqm} mechanism like CoDel for each flow. CoDel aims to keep the queue length under a certain threshold, making sure that the queuing delay was smaller than 5\,ms in the last 100\,ms at least once. Otherwise it drops packets to decrease the queue length. Both of these approaches do not differentiate between flows: They apply the same logic to each flow irrespective of its congestion control and irrespective of the current bandwidth and \gls{rtt}. \cite{bachl_cocoa_2019} showed that this behavior leads to some flows not being able to claim the full bandwidth that they are entitled to. Other flows might achieve the full bandwidth but keep an unnecessary standing queue. The authors then proposed a mechanism that adjusts the queue of each flow based on its congestion control and showed that it works well for several common \glspl{cca}. However, their algorithm has parameters that have to be manually adjusted and it is not guaranteed to work for every \gls{cca} because it makes the assumption that each flow's congestion window follows a zigzag (sawtooth) pattern. Furthermore, their approach is only tailored towards flows that always have data to send while the behavior for application limited flows, that also have idle periods, is not clear. 

Instead of a hand-crafted solution like the existing ones, we argue that instead operators of network hardware should be able to simply specify a reward function, and the \gls{aqm} should then automatically find the right queuing policy based on that reward function using \gls{ml}. We propose \gls{ours}, which achieves these goals. 

To make our work reproducible and to encourage further research, we make the code, figures and trained neural network weights of this work publicly available\footnote{\scriptsize\url{https://github.com/CN-TU/reinforcement-learning-for-per-flow-buffer-sizing}}. 

\section{Related Work}

\cite{bachl_cocoa_2019} uses a hand-crafted algorithm to optimize the queuing behavior for different \glspl{cca}. It is, however, not guaranteed that this algorithm works for all new \glspl{cca} and for all traffic pattern. 

\cite{bless_policy-oriented_2018} develop a steering system for \gls{aqm}, which takes as the input a target utilization (like 95\%) and then tries to adjust the queue so that this target utilization is met. It operates on shared buffers and employs no \gls{ml} and fingerprinting of flows. 

\cite{kim_deep_2019} use Deep Reinforcement Learning to develop an \gls{aqm} mechanism for IoT devices that works on the output queue of a switch.  \cite{bouacida_practical_2019,bisoy_design_2017} use \gls{rl} to build an \gls{aqm} algorithm for shared buffers. \cite{vucevic_reinforcement_2007} uses \gls{rl} as well and focuses specifically on cellular communications. \cite{lin_kemy_2015} use a custom optimization approach that relies on as offline greedy algorithm which finds optimal rules for an \gls{aqm} that fulfills a certain reward function. \cite{shah_sam_2016} build an \gls{aqm} mechanism which utilizes support vector machines to learn an optimal queuing policy for a shared buffer. 

Unlike our proposed approach, these works' aim is not fingerprinting each flow but instead they optimize the overall queue that is shared by flows. In our work we focus on fingerprinting each flow and providing the optimal queuing behavior for each flow individually. This direction has, to our knowledge, not been explored so far. 

Besides \gls{aqm}, Deep \gls{rl} has proven to be successful in several other domains of networking such as Congestion Control \cite{jay_deep_2019,bachl_rax_2019}, Traffic Engineering \cite{xu_experience-driven_2018} and Intelligent Packet Sampling for saving computational resources \cite{bachl_sparseids_2020}. 

\section{Concept}
\label{sec:concept}

\begin{figure}[h]
\includegraphics[width=\columnwidth]{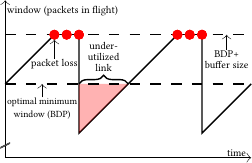}
\caption{If the buffer is too small, loss-based \glspl{cca} cannot fully utilize the link since they send too few data (less than the \gls{bdp}) following the multiplicative decrease that occurs after packet loss.}
\label{fig:tooLittle}
\end{figure}
\begin{figure}[h]
\includegraphics[width=\columnwidth]{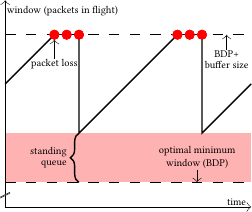}
\caption{If the buffer is too large, loss-based \glspl{cca} keep an unnecessary standing queue (data in flight always more than \gls{bdp}) not required for achieving full link utilization.}
\label{fig:tooMuch}
\end{figure}
\begin{figure}[h]
\includegraphics[width=\columnwidth]{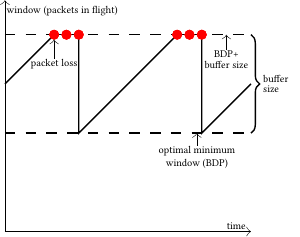}
\caption{For a flow controlled by New Reno, if the buffer is the same size as the \gls{bdp}, no standing queue exists and the full bandwidth is achieved.}
\label{fig:perfect}
\end{figure}

If the buffer for a flow is too small, a flow cannot achieve the full bandwidth as shown in \autoref{fig:tooLittle}. On the other side, if the buffer is too large (\autoref{fig:tooMuch}), the flow can achieve the full bandwidth but a standing queue exists, which causes unnecessary delay.  

For the popular \gls{cca} \textit{New Reno}, the buffer size that is necessary in a fair queuing setting is one \gls{bdp}, meaning $\textit{speed}\times\textit{delay}$ (\autoref{fig:perfect}). This is because its multiplicative decrease factor is $0.5$, meaning that after packet loss, the congestion window is halved. For other \glspl{cca} like \textit{Cubic} \cite{ha_cubic:_2008}, the multiplicative decrease factor is $0.7$, meaning that the congestion window is reduced to 70\% of its previous value upon packet loss. This means that the minimum buffer size required to achieve full throughput for a Cubic flow is $\left(\frac{1}{0.7}-1\right)\times \textit{BDP}$, which is around 43\% of the \gls{bdp} and thus less than for New Reno. That means the minimum required buffer size, among other factors, depends on the \gls{cca} and specifically the \gls{cca}'s multiplicative decrease factor. To reduce packet loss and standing queues we propose an adaptive fair \gls{aqm} that learns the available bandwidth, \gls{rtt} and the underlying \gls{cca} of a flow to adjust the buffer optimally. 

Because the optimal buffer size for many \glspl{cca} depends on the \gls{bdp} and thus on the bandwidth and the delay (\gls{rtt}), for example, a flow only having half the \gls{rtt} of another one only needs half its buffer size to achieve full bandwidth. 

To learn an optimal \gls{aqm} policy, we specify a reward function, which the \gls{ml} based \gls{aqm} should learn to optimize for each flow in a \gls{rl} fashion. A simple and effective reward function is for example:

\begin{align}
\textit{Reward} = \textit{bandwidth}-\alpha\times\textit{queue size}
\label{eq:reward}
\end{align}

In this reward function, the choosable parameter $\alpha$ specifies the tradeoff that is chosen between the bandwidth and the buffer size. With $\alpha$ going to zero, the optimum policy approaches the one in which the buffer size is large enough that a flow never underutilizes the link but at the same time the buffer is never going to be larger than necessary if this doesn't provide a benefit in throughput. Alternatively, the delay could be used instead of the queue size because it is more closely related to \gls{qoe}. However, the delay that is caused by a certain queue length depends on the current link speed. This means that using delay would make the reward function more variable, depending on the current link speed. 

Our \gls{ml} system uses the above reward function to learn the optimal behavior. \gls{ours} outputs the optimal buffer size of which it thinks that it maximizes the reward function when receiving a packet. As the input for this decision we use the following features, which are updated continuously when receiving/sending a packet: 
\begin{itemize}
\item queue size
\item standard deviation of the queue size 
\item maximimum buffer size
\item rate of incoming data
\item rate of outgoing data
\item time since the last packet loss
\end{itemize}
We do not use these features directly because this could result in large, spontaneous variation in the input features. For example, packets might not be received in regular intervals but in bursts. This means that without any smoothening, large instability would occur. Thus, instead we use 10 exponentially weighted averages of each feature with weights of $2^{-4}$, $2^{-5}$, ... , $2^{-13}$. The advantage of using exponentially weighted averages is that they do not occupy any space in memory except for their own values. This is opposed to regular moving averages, which have to keep the entire window of data in memory. For example, for a regular weighted average of window size 1000, the last 1000 data points have to be kept in memory at any time, leading to a large consumption of memory. 

One difficulty is to get an exponentially moving average for the rate of incoming/outgoing packets. Instead we compute the exponentially moving average of the interarrival times of incoming packets and the interdeparture times of outgoing packets. We then invert this number to get the rate. An issue with this approach is that numeric instability and division by zero errors can occur using this approach. Thus, we only use the exponential average with a weight of $2^{-n}$ after at least $n$ packets have arrived. Otherwise we set it to zero. We did this inversion because we saw that using the averages of interarrival/interdeparture times led to slower learning than using the rates. This is because if only the interarrival/interdeparture times are given, the neural network also has to learn the additional step of inverting them, making the learning more time-consuming. 

Using these features gives us a feature vector of $6\times 10 = 60$ features at each point in time. The features are fed into a neural network which has one output: The deemed optimal queue size.  

\section{Implementation} 

We implement \gls{ours} in the network simulator ns-3 \cite{nsnam_ns-3_nodate} and integrate Pytorch's \cite{paszke_pytorch_2019} C++ API into ns-3 for the \gls{dl}. All our code runs on the CPU. 

For the simulations we randomly draw a bandwidth (5 to 25\,Mbit/s), a delay (5 to 25\,ms) and duration (3.75\,ms to 6.25\,ms) and a \gls{cca} (New Reno or BIC \cite{lisong_xu_binary_2004}). We chose these ranges of bandwidths and delays so that the neural network encounters flows with vastly different characteristics and has to learn to deal with them. We use New Reno because it is one of the oldest, most widely deployed \glspl{cca} and BIC because it is similar to Cubic, which is the default in Linux, Windows and most other OSs currently. We cannot use Cubic itself because there is no stable Cubic implementation in ns-3 currently. 

During simulating each flow, at a certain point, we perform an experiment to let the \gls{rl} learn what the optimal buffer size in the current scenario for the current flow would be. As the experiment time, we use a random number between 0 and half of the flow duration of the current flow. The number must be random so that the \gls{rl} learns to take decisions for every stage of a flow: The optimal maximum buffer size is different in the beginning of the flow, when any information about the flow is known yet, than it is after several seconds. The expected behavior is that in the beginning of a flow, the \gls{rl} simply uses the average buffer size that works the best on average for all flows. Then, after more information about the flow is known and it can be fingerprinted effectively, the estimation of the \gls{rl} is supposed to become more fine-grained and specific to the current flow. That means that the \gls{rl} has to learn what the optimal buffer size is for every stage of a flow and that's why the time at which the experiment is performed must be random. Another issue is that if, for example, the experiment is performed just before the end of the flow, the impact of the decision of the \gls{rl} is not clear: If the \gls{rl}, for example, outputs a maximum buffer size of 15 just before the end of the flow, then it has no influence on the flow because the flow ends anyway. That's why we don't perform experiments for the \gls{rl} training in the latter half of a flow.  

For the simulations we use two hosts that are directly connected to each other. We also experimented initially with a switch connecting these hosts but this made the simulation significantly slower. Thus we made the simulation as simple as possible and implemented our \gls{aqm} in front of the bottleneck link that connects the sender to the receiver. For all other buffers we use a FIFO queue with a buffer of 100 packets (the default). 

For the \gls{dl}, we used a fully connected neural network consisting of an input layer, three layers of 256 neurons which each have the leaky ReLU \cite{noauthor_rectifier_2020} function applied and finally an output layer which has an output size of 1. All neural networks we use have this architecture. We chose this number of neurons and layers as recent research shows that generally, overfitting is not a problem, if a neural network is trained long enough \cite{nakkiran_deep_2019}. Thus, it is generally not bad to have more neurons and layers than necessary. That's why we opted for 256 neurons and 3 layers, which we think is quite generous considering that there are only 60 input features and one output. For the learning we chose gradient descent with a learning rate of $0.01$. To save computational resources, how often our \gls{dl} logic runs and outputs the optimal buffer size, is a configurable parameter. The most fine-grained would be to let the \gls{dl} run for each packet that is received. However, this would be a computational waste and thus, by default, we only let the \gls{dl} run for every 10th packet. Other sampling intervals can be configured. Also, an option would be to not let the \gls{dl} run every $n$th packet but every $x$ milliseconds. 

\section{Offline Learning}

First, we implement an \gls{ml} system that is capable of learning an optimal queuing policy in a simulator. The idea is that in an ideal setting training is faster. After training in a simulator, the finished neural network can be deployed in a real production setting. 

The training procedure involves the following steps:
\begin{enumerate}
\item Draw a random sample of 20 bandwidths, delays, \gls{cca} and flow durations.
\item Simulate each flow concurrently, compute the feature vector continuously and let the neural network output its optimal buffer size each time a new packet arrives. 
\item During each flow, at a random time between 0 and half the flow length in seconds, perform the experiment: Fork the simulation process for each flow and continue one simulation with the current buffer size $+1$ packet and one with the current buffer size $-1$ packet until the end of the flow. This procedure is essentially \textit{A/B testing}.  
\item Wait until all flows are finished
\item Check for each flow which of the two versions performed better ($+1$ or $-1$) regarding the reward function.
\item Update the neural network so that it learns to output the better buffer size when being fed the inputs as they were at the time at which the experimentation started. Specifically, we compute the \gls{mae} between the output of the neural network and the desired output (the one that performed better in the experiment) for the whole batch of 20 results.
\item Start the next iteration. 
\end{enumerate}

During deployment, the trained neural network is used but the experiment step is skipped. We chose to run 20 simulations concurrently since our computers have 40 CPUs and every simulation is split in two at the time of the experiment when each simulation process is forked. Thus with 20 simulations we fully utilize the 20 CPUs. This implies that the batch size used for the \gls{dl} is 20. 

\subsection{$\alpha=0.01$}

\begin{figure}[h]
\includegraphics[width=\columnwidth]{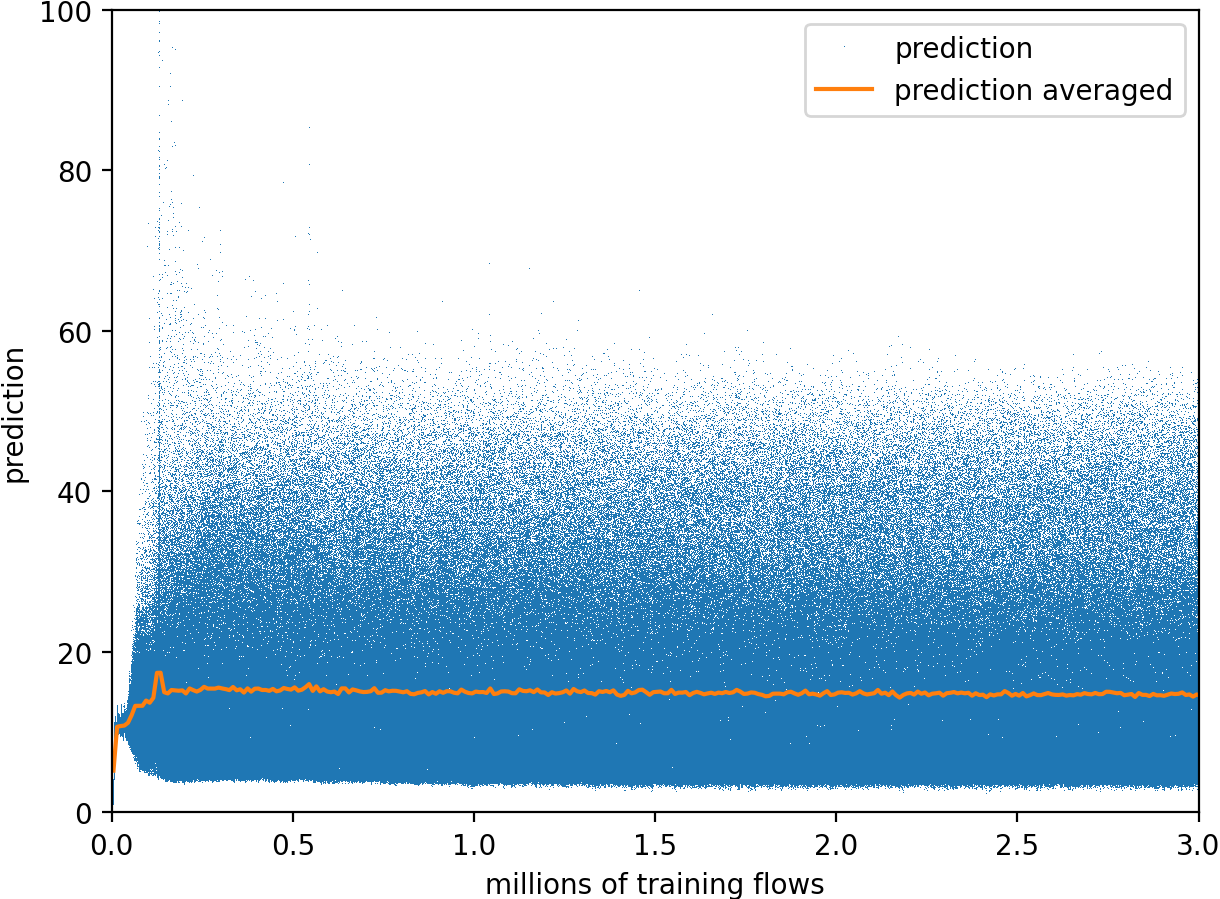}
\caption{The prediction of the optimal buffer by the neural network during offline training with the tradeoff $\alpha=0.01$. The large variance in predicted optimal buffer sizes is the intended result as this means that different optimal buffer sizes are output for different types of flows, which means that the neural network can differentiate different flows.}
\label{fig:offlineTraining}
\end{figure}

When performing offline learning with the tradeoff parameter $\alpha=0.01$ (\autoref{fig:offlineTraining}), the neural network first starts with a prediction that is close to 0 (due to the initialization of neural network weights). During the first couple of thousand training flows, it increases its output and approaches what appears to be the buffer size that works best on average (slightly higher than 15). Then the neural network gradually starts understanding and differentiating different flows and learns custom policies for them. After around 250000 flows, the output appears to be stable and doesn't change anymore. There's one noticeable outlier at around 120000 flows. This appears to be some instability that can commonly occur in the early stages of neural network training but which has no significance for the training process since it resumes normally after the outlier. 

\begin{figure}[h]
\includegraphics[width=\columnwidth]{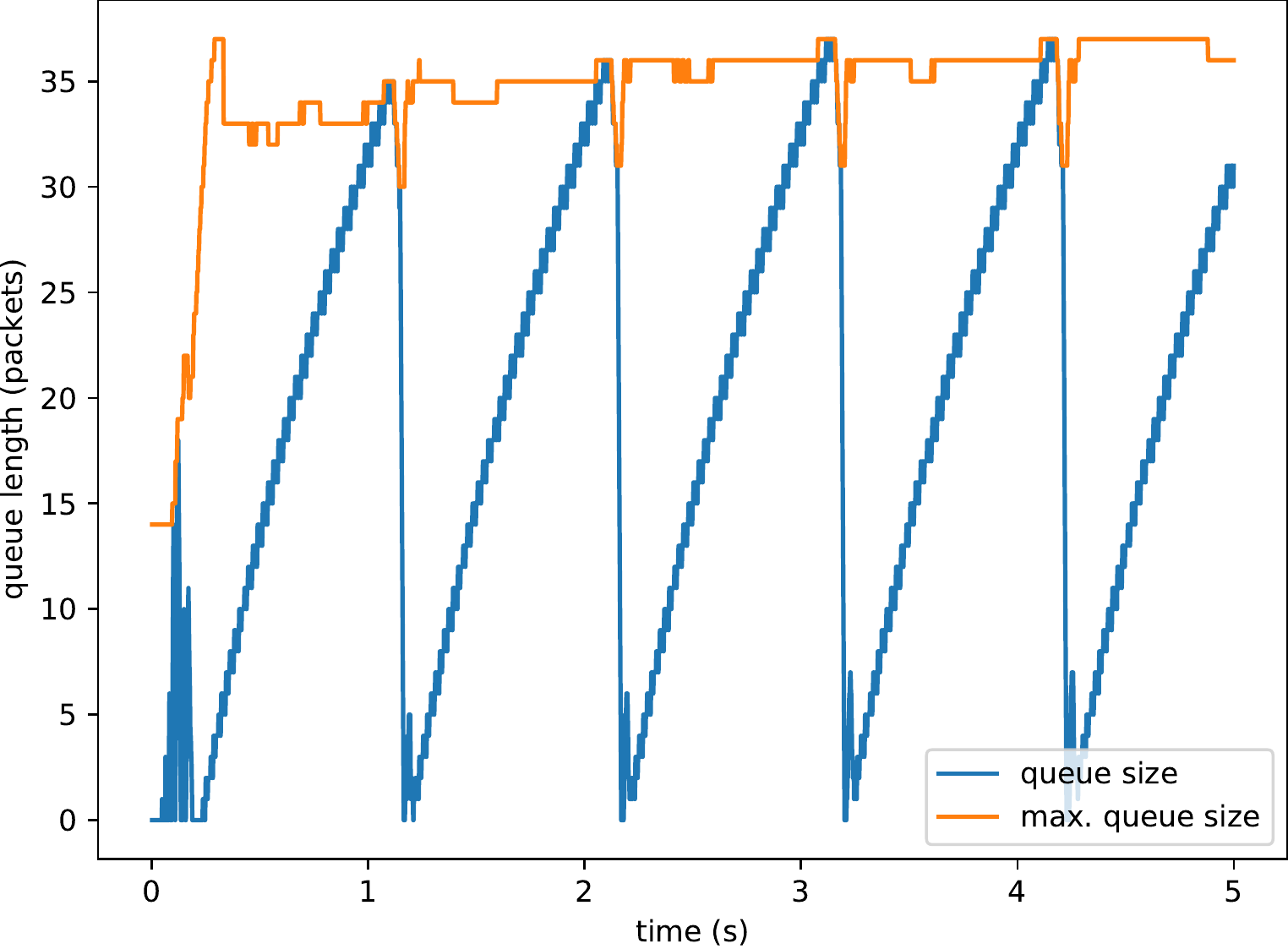}
\caption{Showing the maximum queue length and queue length of a New Reno flow controlled by \gls{ours} (offline training, $\alpha=0.01$) at the bottleneck with a 25\,Mbit/s link and a delay of 15\,ms. The queue never becomes empty and also never forms a standing queue.}
\label{fig:exampleRenoLargeBw}
\end{figure}

\autoref{fig:exampleRenoLargeBw} shows a New Reno flow at a link with a bottleneck bandwidth of 25\,Mbit/s and a delay of 15\,ms controlled by \gls{ours} after it is fully trained. In the beginning, the maximum queue length that \gls{ours} allows is around 15 packets. After less than a second, when the \gls{dl} learns to understand the flow's congestion control, the bottleneck delay and bandwidth, it understands that a larger maximum queue is needed and raises its prediction to around 35 packets. It stays at this value until the end. The queue never becomes empty and no standing queue is formed at any point, which means that \gls{ours} achieved its goal.

\begin{figure}[h]
\includegraphics[width=\columnwidth]{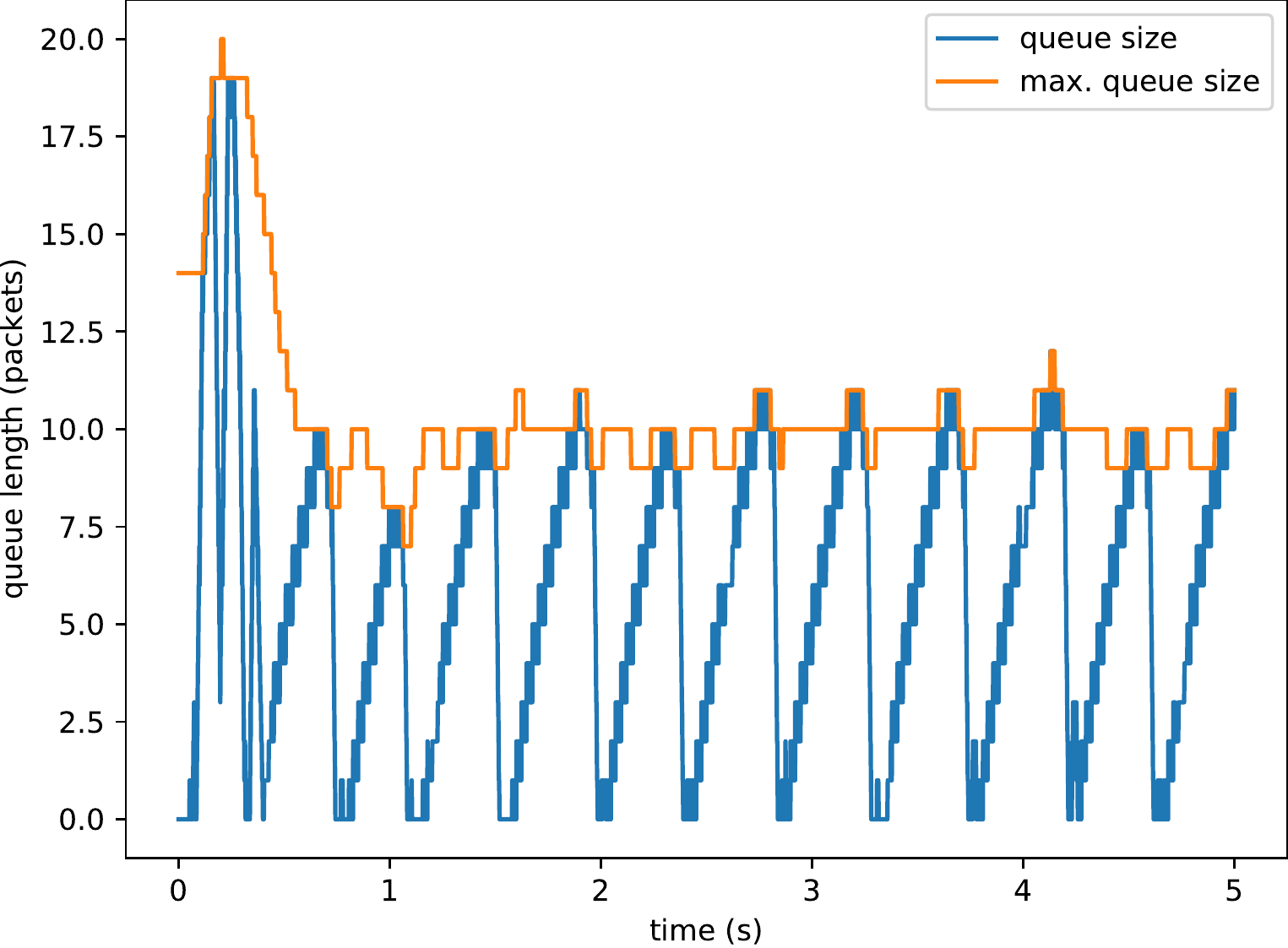}
\caption{Showing the maximum queue length and queue length of a New Reno flow controlled by \gls{ours} (offline training, $\alpha=0.01$) at the bottleneck with a 6\,Mbit/s link and a delay of 15\,ms. The queue never becomes empty and also never forms a standing queue.}
\label{fig:exampleRenoSmallBw}
\end{figure}

\autoref{fig:exampleRenoSmallBw} shows a New Reno flow at a link with a bottleneck bandwidth of 6\,Mbit/s and a delay of 15\,ms controlled by \gls{ours}. In the beginning, the maximum queue length that \gls{ours} allows is around 15 packets, like for \autoref{fig:exampleRenoLargeBw}. Then, however, it behaves differently than the flow before on the link with a larger bandwidth: \gls{ours} learns that the bottleneck link has a smaller bandwidth and thus, a smaller buffer is sufficient. Thus, after less than a second, it learns that the optimal maximum queue size is around 10 packets. Then it keeps that buffer size until the end of the flow. Again, the queue never becomes empty for a prolonged time and no standing queue forms. 

\begin{figure}[h]
\includegraphics[width=\columnwidth]{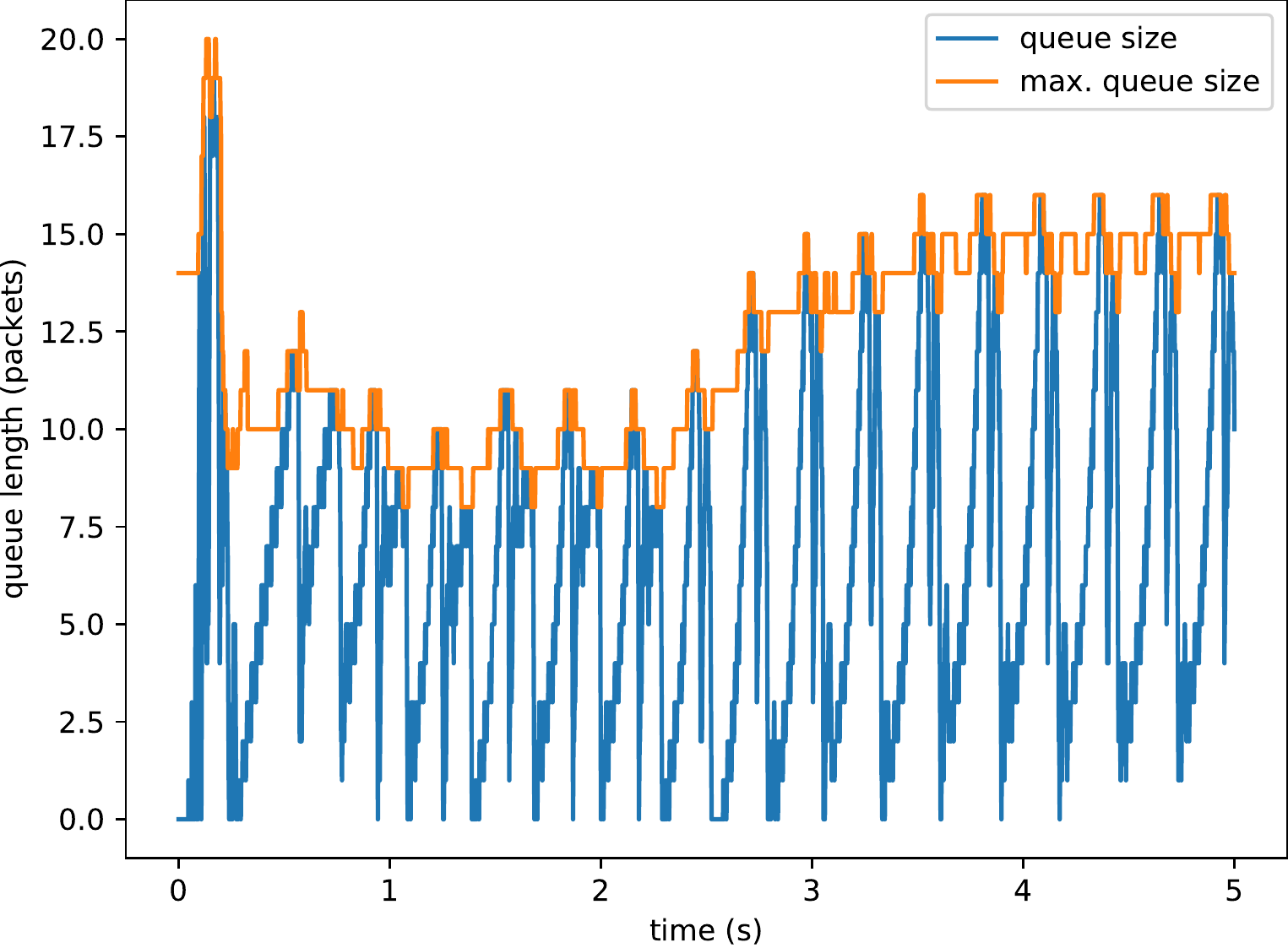}
\caption{Showing the maximum queue length and queue length of a BIC flow controlled by \gls{ours} (offline training, $\alpha=0.01$) at the bottleneck with a 25\,Mbit/s link and a delay of 15\,ms. The queue never becomes empty and also never forms a standing queue.}
\label{fig:exampleBicLargeBw}
\end{figure}

\autoref{fig:exampleBicLargeBw} shows that \gls{ours} learned that the flow uses the BIC \gls{cca} and thus needs a smaller buffer: For the same scenario with New Reno (\autoref{fig:exampleRenoLargeBw}) \gls{ours} output a queue length of around 35. For the same scenario with a BIC controlled flow, it chooses the optimal maximum queue length as around 12 packets. This demonstrates that \gls{ours} learned to distinguish different congestion controls only by looking at the pattern of a flow. 

\begin{table}[h]
\caption{Correlation between bandwidth/delay of the link and output buffer size for a tradeoff of $0.1$ with offline learning. We vary bandwidth between 5 and 25\,Mbit/s and delay between 5 and 25\,ms for both \glspl{cca}. When bandwidth is varied, delay is kept at its mean (15\,ms) and vice versa.} \label{tab:corrOfflineSmallAlpha}
\centering
\begin{tabular}{lrr} \toprule
\gls{cca} & bandwidth & delay \\ \midrule
New Reno & 98.3\% & 99.1\% \\
BIC & 80.2\% & 90.2\% \\
\bottomrule
\end{tabular}
\end{table}

To have a quantifiable measure of the success of \gls{ours}, we compute the correlation between a change of bandwidth/delay and the maximum queue size output by the neural network. If the neural network learned ideally, the correlation obtained would be 1 (100\%) as, as mentioned in \autoref{sec:concept}, with a larger delay/bandwidth the \gls{bdp} is larger and a larger buffer is necessary. \autoref{tab:corrOfflineSmallAlpha} shows that the for New Reno the correlation is very high while for BIC it is high but not as good as for New Reno. We attribute this to the fact that for BIC the buffer that is required is generally smaller (only a couple of packets) and thus fluctuations in the training process are more pronounced. 

The correlations imply that, when increasing the bandwidth, the neural network outputs a larger maximum queue size, because a larger buffer is needed for New Reno and BIC, when the bandwidth is larger. However, learning the relationship is not difficult: In fact, the neural network would simply need to learn to use the identity function to map the input feature that encodes the incoming or outgoing data rate to the output (the output maximum buffer size). 

The results also show that increasing the delay increases the output (maximum buffer size) of the neural network. This is more interesting than for the bandwidth since no input feature directly encodes the delay. Thus, the neural network must observe the pattern of the \gls{cca}, estimate the \gls{rtt} and output the suitable maximum queue size. The right column of \autoref{tab:corrOfflineSmallAlpha} shows that there is clearly a linear relationship between the delay and the output maximum queue size as the correlation is close to 1 (or 100\%). 

\begin{table}[h]
\caption{The average maximum queue length (in packets) output by the neural network after offline learning (tradeoff $0.01$) when averaging over flows with varying bandwidth/delay like in \autoref{tab:corrOfflineSmallAlpha}.} \label{tab:avgOfflineSmallAlpha}
\centering
\begin{tabular}{lr} \toprule
\gls{cca} & avg.~max.~queue length \\ \midrule
New Reno & 22 \\
BIC & 8 \\
\bottomrule
\end{tabular}
\end{table}

\autoref{tab:avgOfflineSmallAlpha} shows that the average output maximum queue size is more than double for New Reno than for BIC. This shows that the neural network also learned to distinguish between these two \glspl{cca}. Since there are no features that directly indicate the \gls{cca}, the neural network must have learned these by combining the other features and observing distinct behavior for New Reno and BIC. This is expected, since BIC has a different multiplicative decrease factor than New Reno ($0.7$ vs $0.5$) the neural network should learn to output different maximum queue sizes for these two. Specifically, we would expect BIC to need a smaller buffer as its decrease is smaller, thus requiring a smaller buffer. 

\begin{figure*}[h]
\centering
\subfloat[New Reno, varying bandwidth.\label{fig:SmallAlphaNewRenoBandwidth}
]{
\includegraphics[width=0.98\columnwidth]{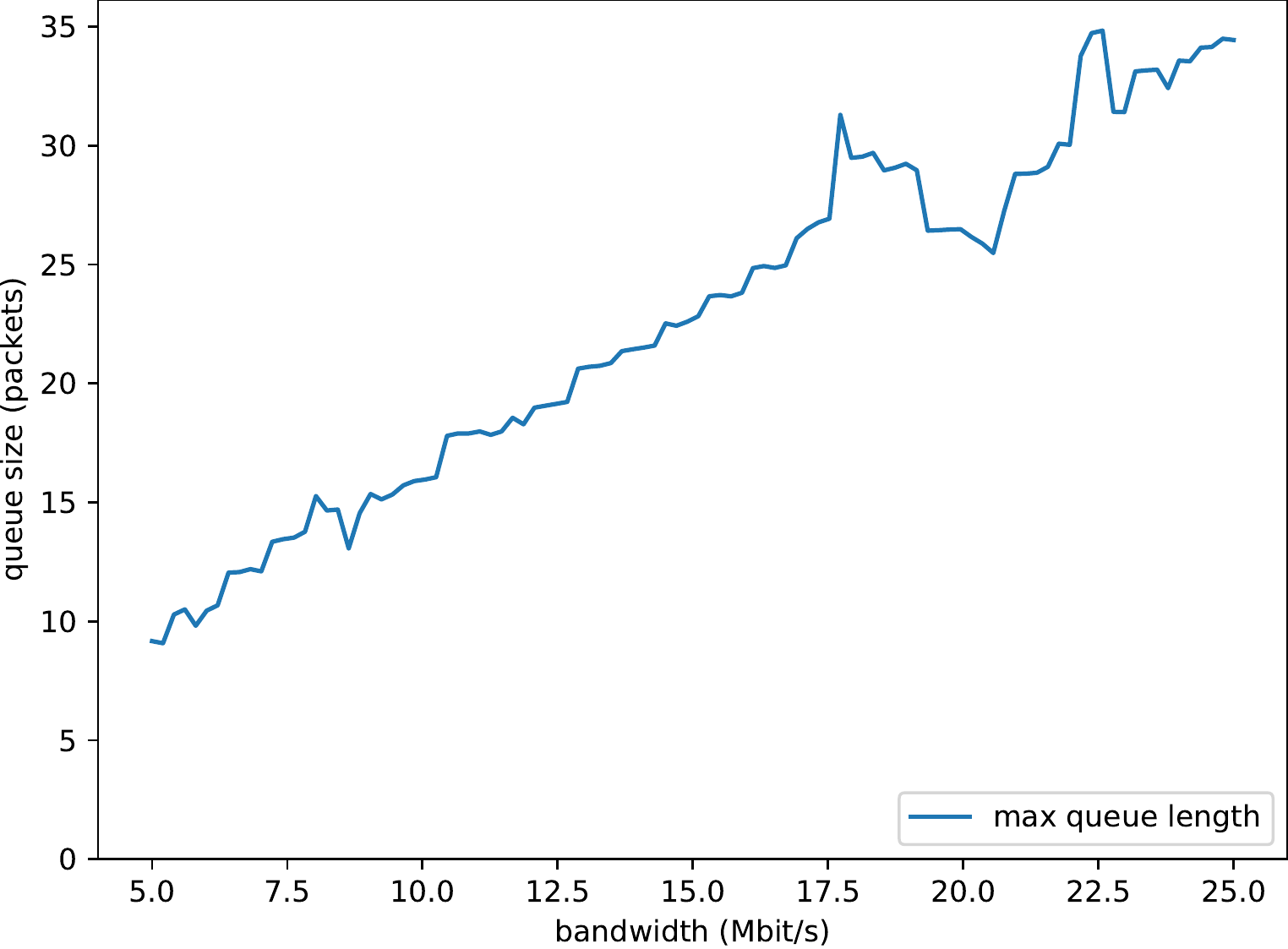}
}{}
\subfloat[New Reno, varying delay.\label{fig:SmallAlphaNewRenoDelay}
]{
\includegraphics[width=0.98\columnwidth]{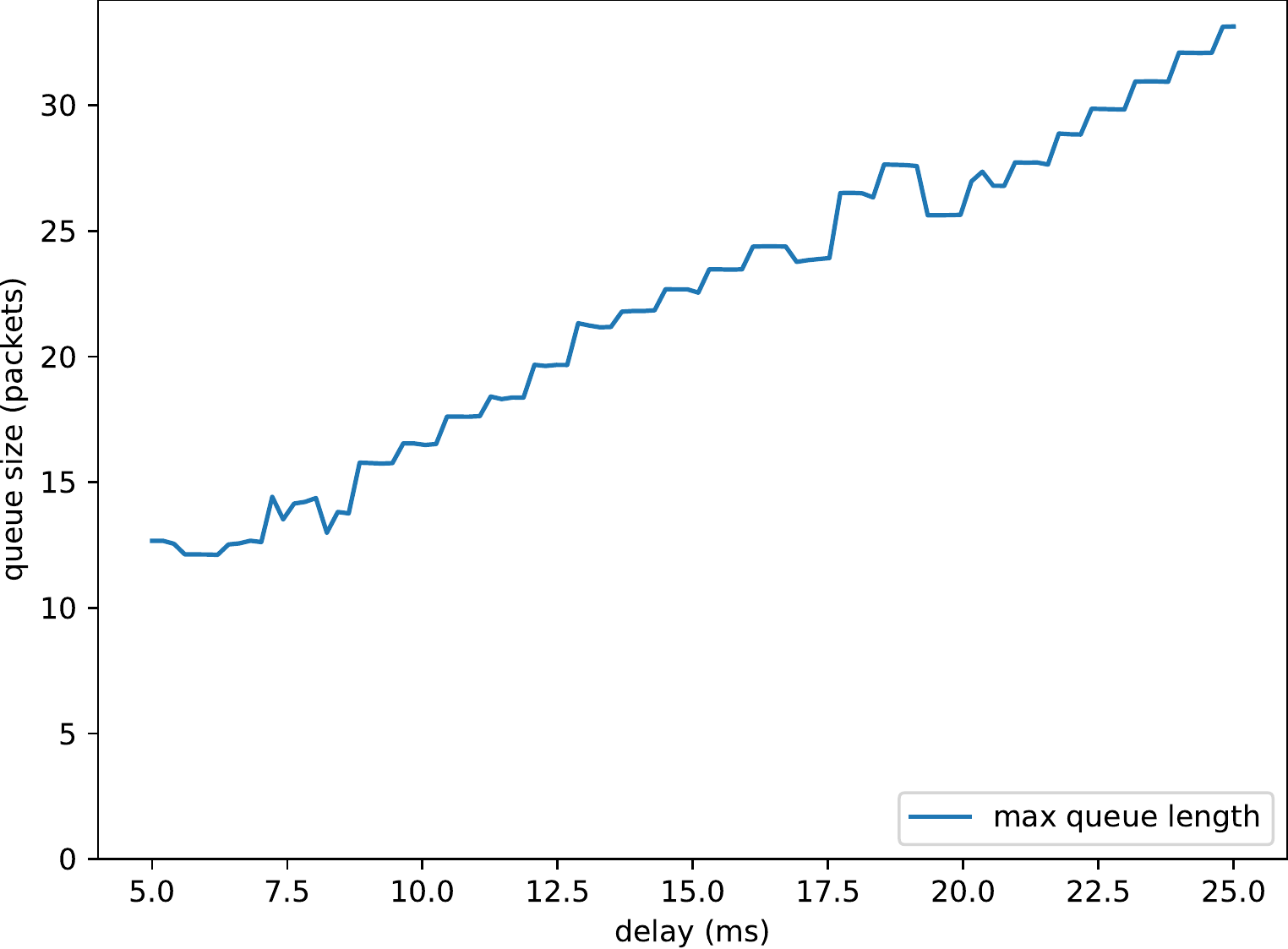}
}{}
\subfloat[BIC, varying bandwidth\label{fig:SmallAlphaBicBandwidth}
]{
\includegraphics[width=0.98\columnwidth]{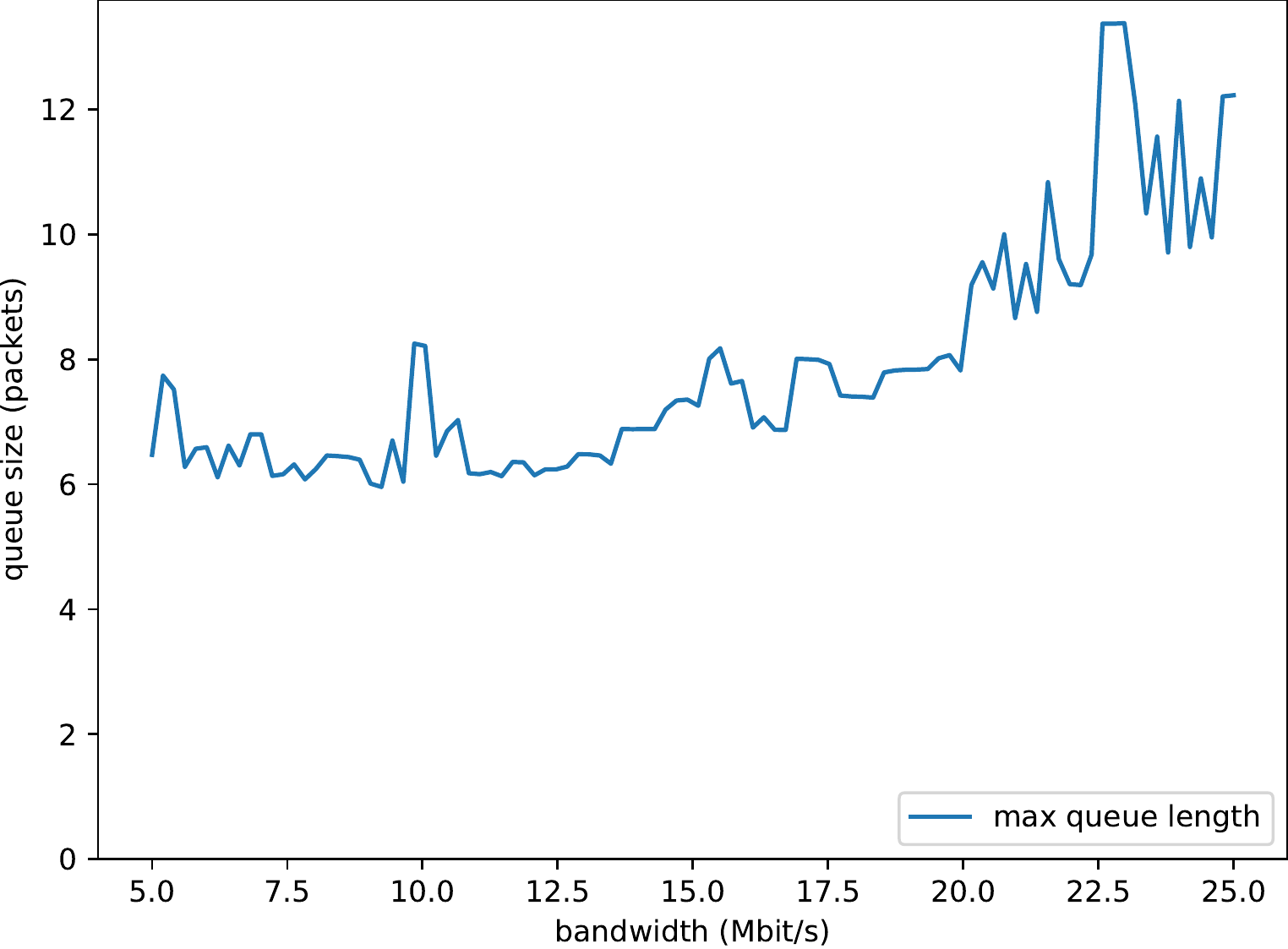}
}{}
\subfloat[BIC, varying delay\label{fig:SmallAlphaBicDelay}
]{
\includegraphics[width=0.98\columnwidth]{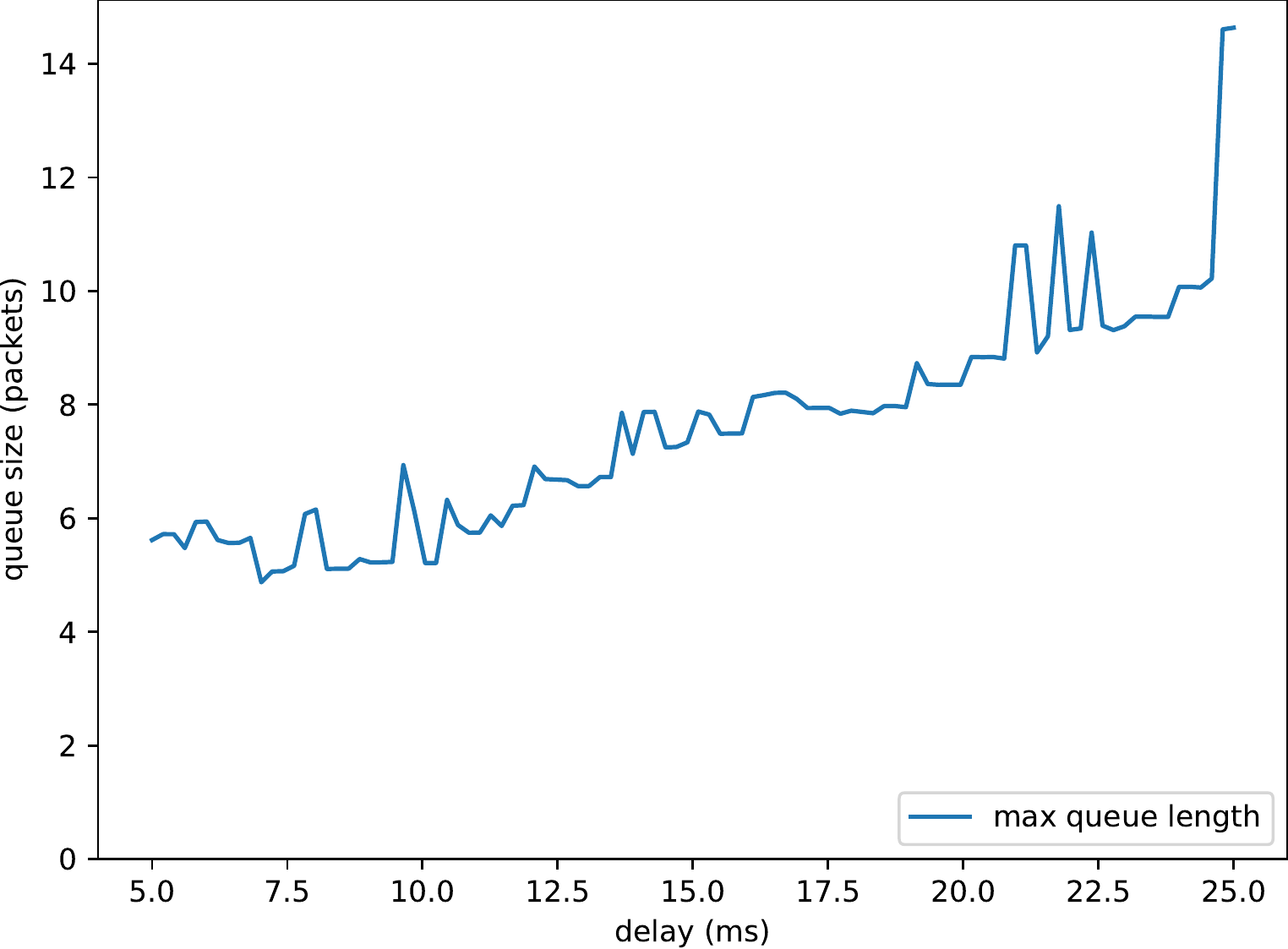}
}{}
\caption{Change of the buffer size after offline learning when varying bandwidth/delay/\gls{cca}. When bandwidth is varied, delay is kept in the middle of its range, at 15\,ms. Vice versa, when delay is varied, bandwidth is kept in the middle of its range, at 15\,Mbit/s.}
\label{fig:offlineSmallAlpha}
\end{figure*}

\autoref{fig:offlineSmallAlpha} depicts how the output maximum buffer size changes when delay/bandwidth and the \glspl{cca} are varied. It shows the linear association between delay/bandwidth and output maximum buffer size as well as the generally lower buffer size that is output for BIC flows. There are some fluctuations, which are especially apparent for BIC which we attribute to the fact that for small buffer sizes, fluctuations in the training process are more pronounced. 

\subsection{$\alpha=10$}

\begin{figure}[h]
\includegraphics[width=\columnwidth]{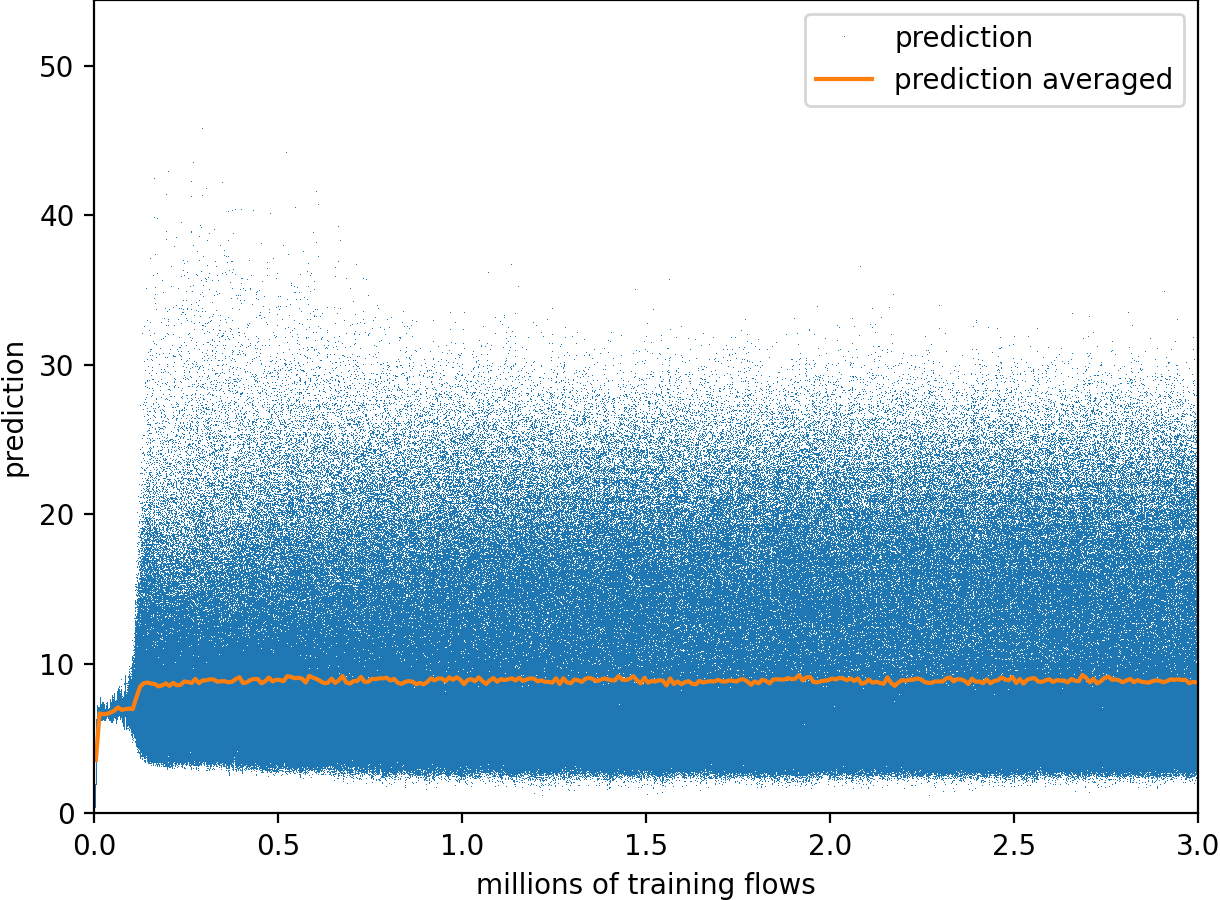}
\caption{The prediction of the optimal buffer by the neural network during offline training with the tradeoff $\alpha=10$.}
\label{fig:offlineTrainingLargeAlpha}
\end{figure}

\autoref{fig:offlineTrainingLargeAlpha} shows that also for the larger $\alpha$, the neural network converges quickly after a couple of 10000 flows. 

\begin{figure}[h]
\includegraphics[width=\columnwidth]{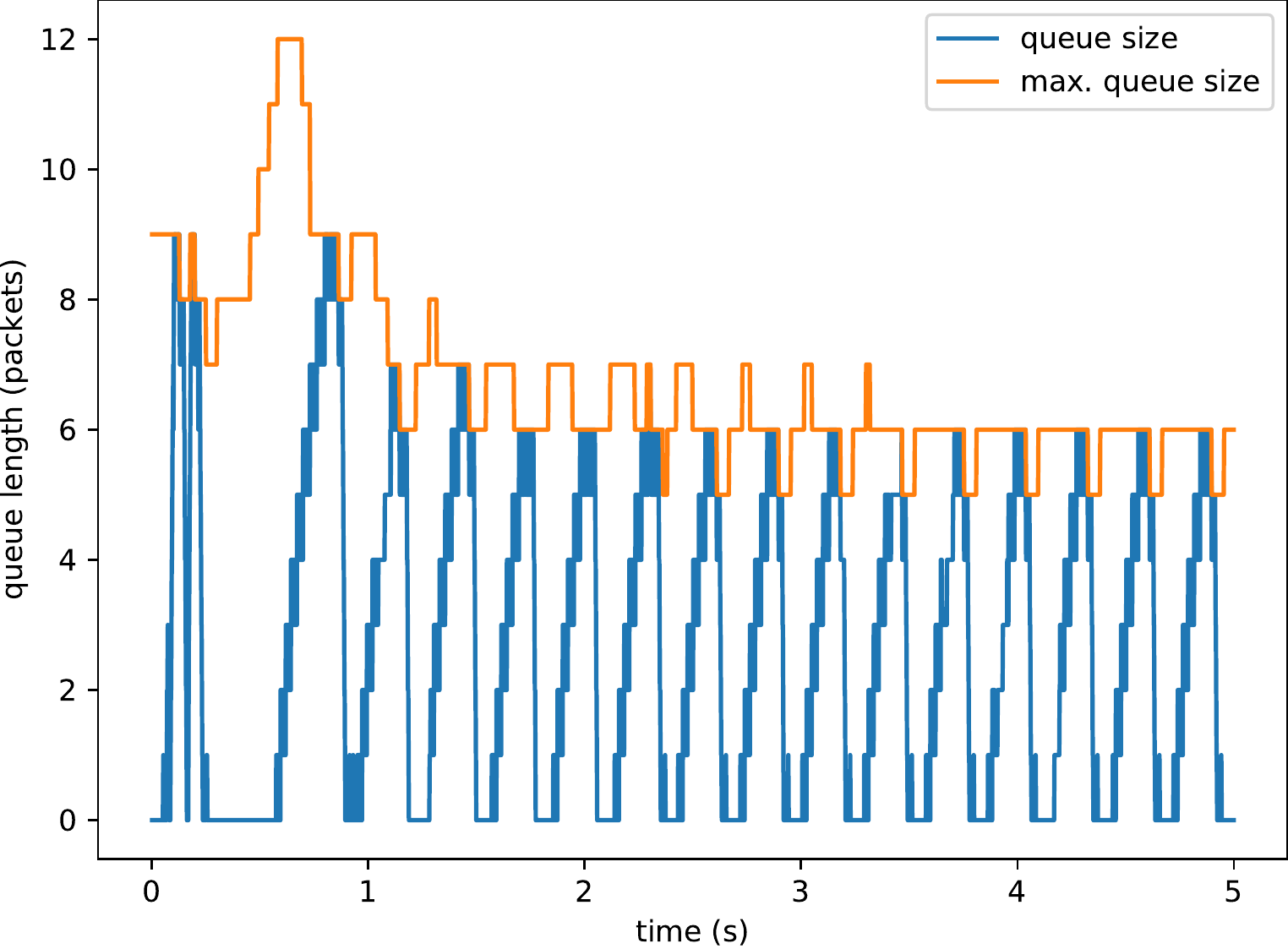}
\caption{Showing the maximum queue length and queue length of a New Reno flow controlled by \gls{ours} (offline training, $\alpha=10$) at the bottleneck with a 6\,Mbit/s link and a delay of 15\,ms. The queue occasionally becomes empty which is intended as with the higher tradeoff $\alpha$, \gls{ours} tries to favor a small queue more than high throughput.}
\label{fig:exampleRenoSmallBwLargeAlpha}
\end{figure}

\autoref{fig:exampleRenoSmallBwLargeAlpha} shows a flow under the same conditions like the one in \autoref{fig:exampleRenoSmallBw}. The flow in \autoref{fig:exampleRenoSmallBwLargeAlpha} however, has a smaller buffer and sometimes the queue becomes empty, which never happens for the flow in \autoref{fig:exampleRenoSmallBw}. This shows that the parameter $\alpha$ has a clear influence on the learning process and that with a larger $\alpha$, throughput is sacrificed to keep the queue smaller. 

\begin{table}[h]
\caption{Correlation between bandwidth/delay of the link and output buffer size for a tradeoff of $10$ with offline learning. We vary bandwidth between 5 and 25\,Mbit/s and delay between 5 and 25\,ms for both \glspl{cca}. When bandwidth is varied, delay is kept at its mean (15\,ms) and vice versa.} \label{tab:corrOfflineLargeAlpha}
\centering
\begin{tabular}{lrr} \toprule
\gls{cca} & bandwidth & delay \\ \midrule
New Reno & 99.3\% & 96.5\% \\
BIC & 75.4\% & 62\% \\
\bottomrule
\end{tabular}
\end{table}

\begin{table}[h]
\caption{The average maximum queue length (in packets) output by the neural network after offline learning (tradeoff $10$) when averaging over flows with varying bandwidth/delay like in \autoref{tab:corrOfflineLargeAlpha}.} \label{tab:avgOfflineLargeAlpha}
\centering
\begin{tabular}{lr} \toprule
\gls{cca} & avg.~max.~queue length \\ \midrule
New Reno & 13 \\
BIC & 5 \\
\bottomrule
\end{tabular}
\end{table}

Results in \autoref{tab:corrOfflineLargeAlpha} and \autoref{tab:corrOfflineLargeAlpha} show that for a larger $\alpha$ \gls{ours} still works as expected and that the average output maximum queue size is smaller, as expected. 

\section{Online Learning}

In contrast to offline learning, online learning enables to train in deployment. This makes it possible to adapt the behavior slowly over time (over a time span of months or years), for example if new \glspl{cca} emerge. Furthermore, being able to training with real flows is more realistic than training only with simulated flows in a simulation. 

The major change is that for online learning, A/B testing, like for the offline learning, is not possible: In the real world, each time the buffer size is updated for a flow, this is a definite decision. It is not possible to artificially ``split'' a flow like in the simulator and check if a larger or a smaller buffer would have been a better choice. Another analogy for this is that for online learning, we have to unveil what happened to Schrödinger's cat (if it is dead or not), when we launch the experiment, which buffer size is better. For the offline learning, on the other side, we can try out both versions in the simulator and continue one reality with the cat being dead (maximum buffer size $-1$) and the other one with the cat being alive (maximum buffer size $+1$). 

To accommodate for this difference from offline learning, for online training we need two neural networks: 
\begin{itemize}
\item An actor network, which outputs the optimal buffer size at each time step.
\item A critic network, which outputs the reward that it predicts is going to be achieved when keeping the current buffer size until the end of the flow. 
\end{itemize}
Both of these neural networks have the same number of layers and neurons. 

The training procedure involves the following steps:
\begin{enumerate}
\item Draw a random sample of 40 bandwidths, delays, \gls{cca} and flow durations.
\item Simulate each flow concurrently, compute the feature vector continuously and let the neural network output its optimal buffer size each time a new packet arrives. 
\item During each flow, at a random time between 0 and the flow length in seconds divided by 2, perform the experiment: Continue the simulation either with the current buffer size incremented by 1 or decremented by 1 until the end of the flow.
\item Wait until all flows are finished
\item Check for each flow if its decision was better than the expectation of the critic network.
\item Update the actor neural network to output the better buffer size when being fed the inputs as they were at the time at which the experimentation started. For example, if the experiment was with buffer size $-1$ but this yielded worse results than expected, this implies that the buffer size $+1$ would have been better. As for offline training we use the \gls{mae}, averaging over the results of the whole batch of 40 flows. 
\item Update the critic neural network by using the input vector that was recorded at the time at which the experiment started and the reward that was achieved as the label. Specifically, \gls{mse} is used as the loss function, averaging over the whole batch of 40 flows. 
\item Start the next iteration. 
\end{enumerate}

For online learning we run 40 simulations concurrently since we have 40 CPUs and online learning does not split each simulation into two as it happens for offline learning. The batch size for the \gls{dl} is thus 40. 

\subsection{$\alpha=10$}

\begin{figure}[h]
\includegraphics[width=\columnwidth]{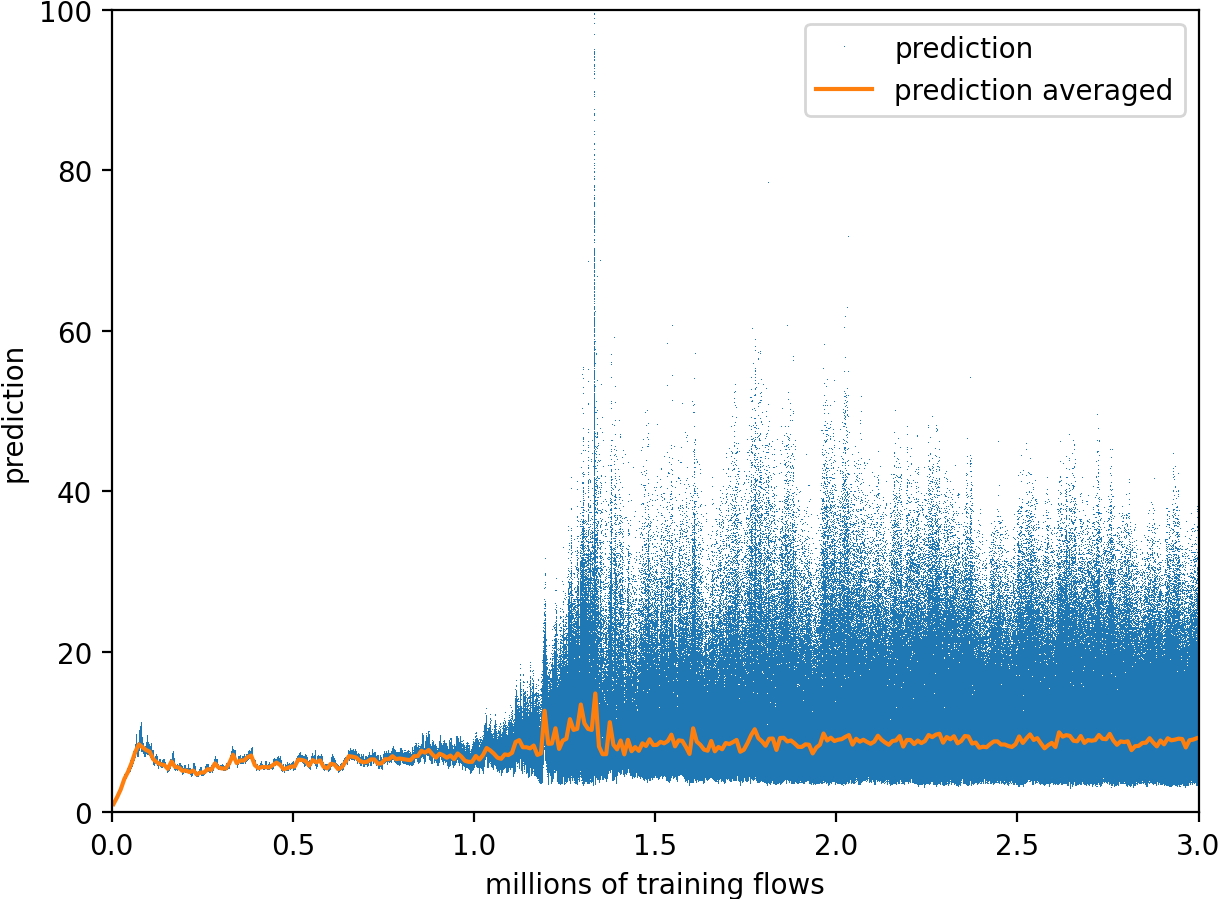}
\caption{The prediction of the optimal buffer by the neural network during online training with the tradeoff $\alpha=10$.}
\label{fig:onlineTraining}
\end{figure}

\autoref{fig:onlineTraining} shows that it takes significantly more network flows for training to converge. This is not surprising since for online training there are two neural networks involved and there is inherently more noise: For the offline training, we can perform A/B testing and thus the reward given to the neural network is always correct. On the other side, for online learning, the reward depends also on the output of the critic network and the critic network doesn't have to be 100\% correct when it outputs the reward that it expected. Thus, the more noisy reward for online training makes it converge more slowly. Specifically, this means that first, the neural network learns that a buffer size of around 10 is optimal on average for all flows and only very slowly (when compared to the offline learning) it learns to fingerprint and differentiate the different flows with different conditions such as different link speed, different delay and different \gls{cca}. 

\begin{table}[h]
\caption{Correlation between bandwidth/delay of the link and output buffer size for a tradeoff of $10$ with online learning. We vary bandwidth between 5 and 25\,Mbit/s and delay between 5 and 25\,ms for both \glspl{cca}. When bandwidth is varied, delay is kept at its mean (15\,ms) and vice versa.} \label{tab:corrOnlineLargeAlpha}
\centering
\begin{tabular}{lrr} \toprule
\gls{cca} & bandwidth & delay \\ \midrule
New Reno & 86.1\% & 87.7\% \\
BIC & 72.9\% & 73.6\% \\
\bottomrule
\end{tabular}
\end{table}

\begin{table}[h]
\caption{The average maximum queue length (in packets) output by the neural network after online learning (tradeoff $\alpha=10$) when averaging over flows with varying bandwidth/delay like in \autoref{tab:corrOnlineLargeAlpha}.} \label{tab:avgOnlineLargeAlpha}
\centering
\begin{tabular}{lr} \toprule
\gls{cca} & avg.~max.~queue length \\ \midrule
New Reno & 14 \\
BIC & 8 \\
\bottomrule
\end{tabular}
\end{table}

\autoref{tab:corrOnlineLargeAlpha} and \autoref{tab:avgOnlineLargeAlpha} show that online learning also learns the expected mapping between \gls{cca}, bandwidth and delay to buffer size, albeit training takes longer and results are more noisy than for online learning as can be seen in the lower correlations when comparing \autoref{tab:avgOfflineLargeAlpha} with \autoref{tab:avgOnlineLargeAlpha}

\section{Comparing with other queue managers}

In this section we compare the behavior of \gls{ours} to the behavior of other popular fair queue discs: \begin{itemize}
\item fq \cite{dumazet_pkt_sched:_2013}, which is a common fair queue disc for Linux which has a drop tail buffer of fixed size for each flow (Fifo) with two different buffer sizes: 100 and 1000, which both are common default buffer sizes. 
\item FqCoDel \cite{taht_flow_2018}, which uses fq but doesn't manage each queue with a drop tail buffer but instead with the CoDel algorithm. 
\end{itemize}

\begin{figure}[h]
\includegraphics[width=\columnwidth]{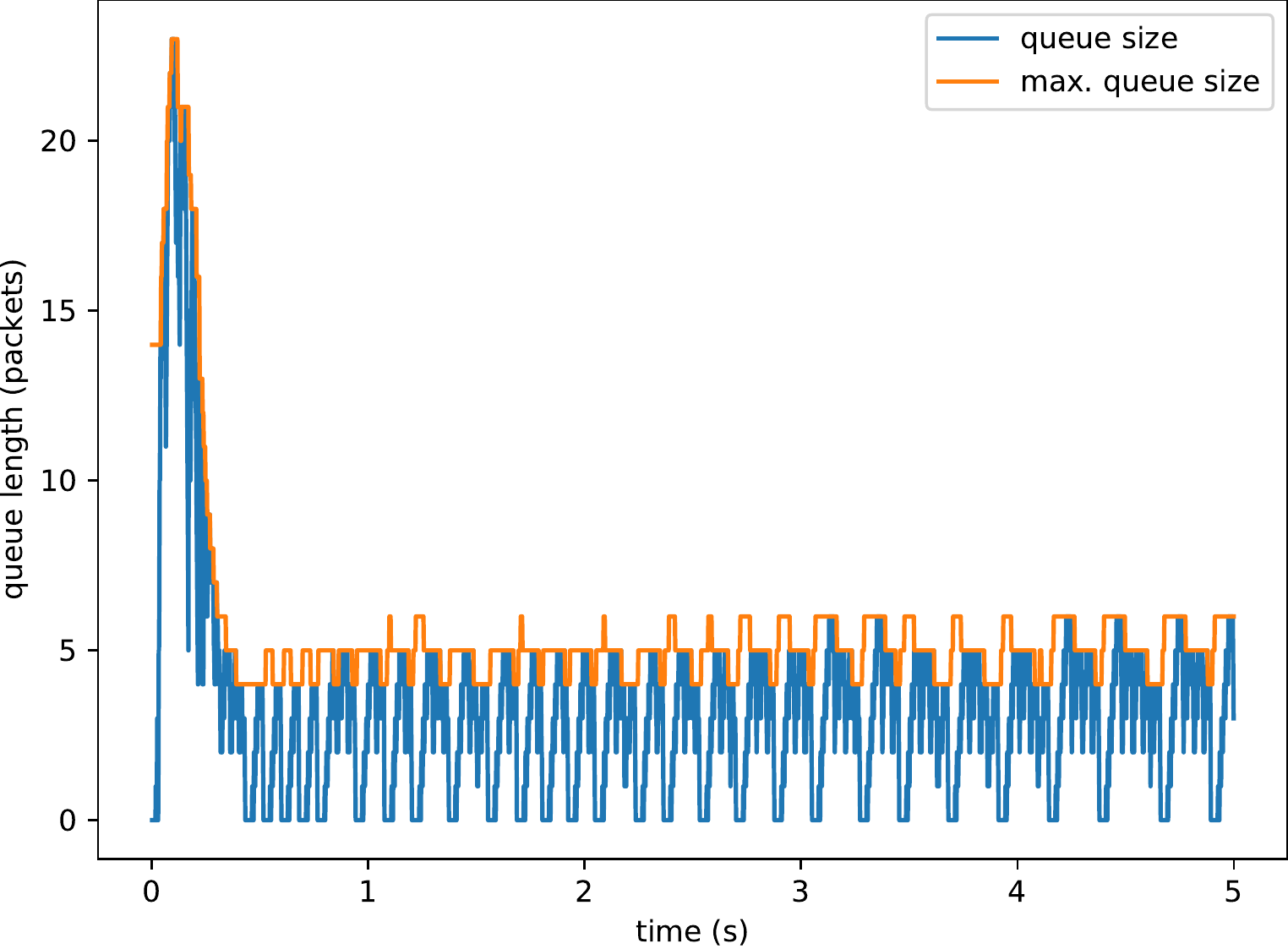}
\caption{Showing the maximum queue length and queue length of a BIC flow controlled by \gls{ours} (offline training, $\alpha=0.01$) at the bottleneck with a 15\,Mbit/s link and a delay of 5\,ms. The queue never becomes empty and also never forms a standing queue.}
\label{fig:exampleBicOursNice}
\end{figure}

\begin{figure}[h]
\includegraphics[width=\columnwidth]{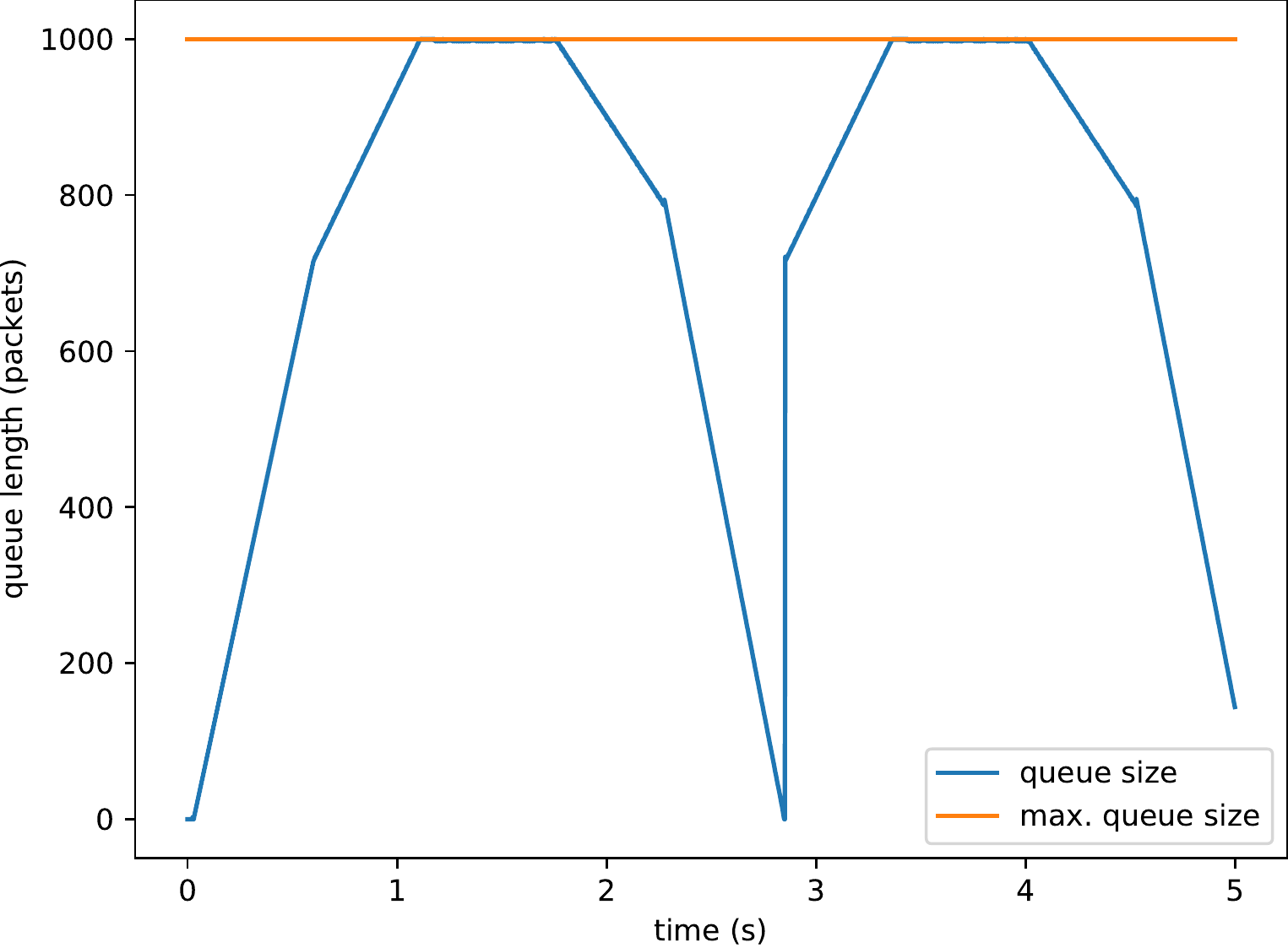}
\caption{Showing the maximum queue length and queue length of a Bic flow controlled by fq with a max queue of 1000 packets at the bottleneck with a 15\,Mbit/s link and a delay of 5\,ms.}
\label{fig:exampleBicfq1000}
\end{figure}

\begin{figure}[h]
\includegraphics[width=\columnwidth]{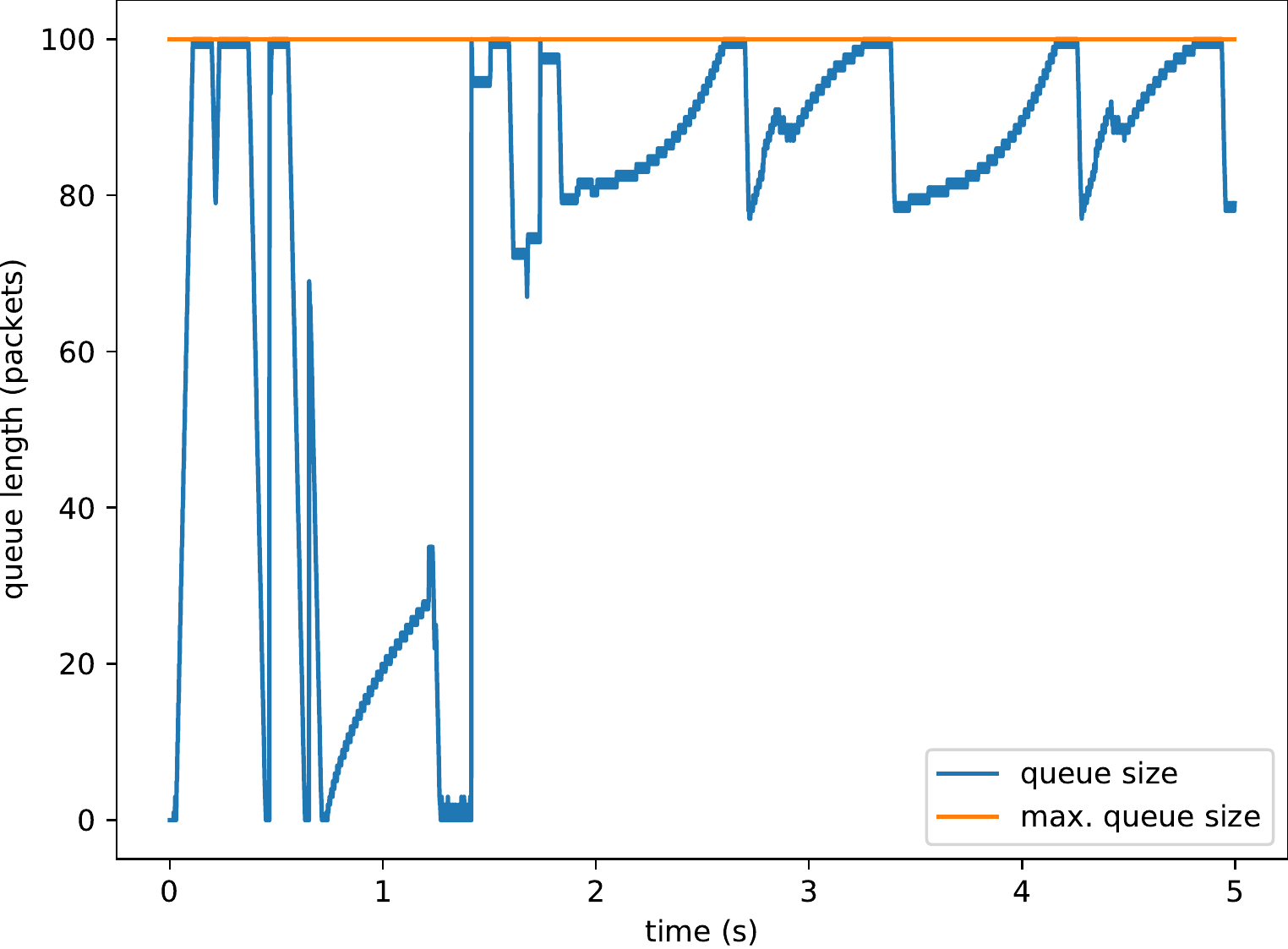}
\caption{Showing the maximum queue length and queue length of a Bic flow controlled by fq with a max queue of 1000 packets at the bottleneck with a 15\,Mbit/s link and a delay of 5\,ms. A large standing queue of about 80 packets is kept starting from second 2.}
\label{fig:exampleBicfq100}
\end{figure}

\begin{figure}[h]
\includegraphics[width=\columnwidth]{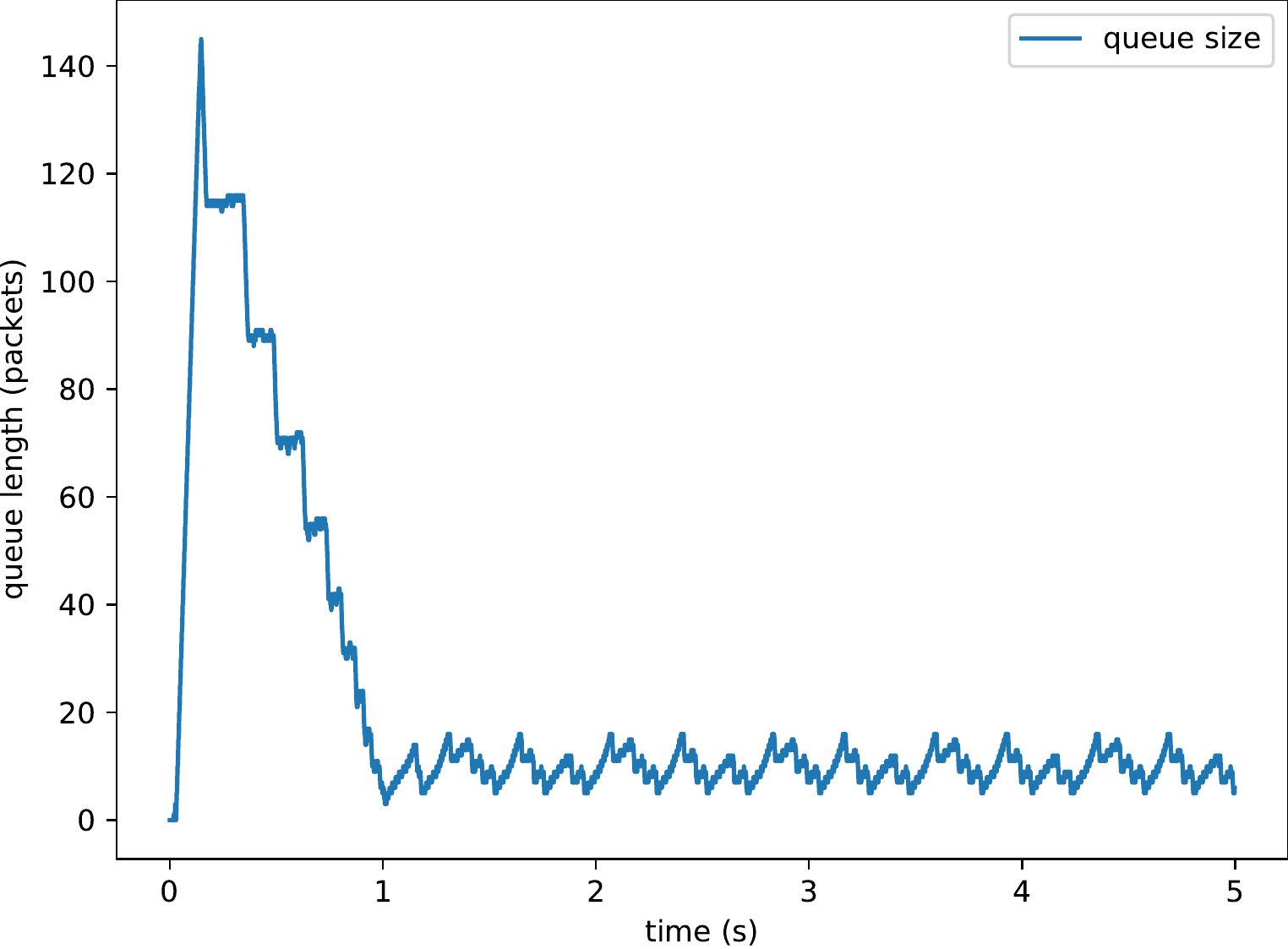}
\caption{Showing the maximum queue length and queue length of a Bic flow controlled by FqCoDel at the bottleneck with a 15\,Mbit/s link and a delay of 5\,ms.}
\label{fig:exampleBicFqCoDel}
\end{figure}

Comparing the behavior of \gls{ours} (\autoref{fig:exampleBicOursNice}) with the behavior of fq (\autoref{fig:exampleBicfq1000} and \autoref{fig:exampleBicfq100}) for an example flow shows that while fq keeps a very large standing queue, \gls{ours} keeps the queue at the minimum so that it is possible to achieve maximum throughput and minimum queue size according to our reward function (\autoref{eq:reward}). The comparison with FqCoDel (\autoref{fig:exampleBicFqCoDel}) shows that FqCoDel generally avoids a large standing queue but results in a very large spike in queued packets at the beginning of the flow. Thus, while CoDel was designed to avoid ``bufferbloat'' it actually induces a very large standing queue during the slow start phase of a flow. 

\begin{table}
\caption{Comparing the average throughput and maximum/average queue size over a large range of network conditions for New Reno and Bic. All results are the average of 100 experiments with delay ranging from 5 to 25\,ms (with bandwidth fixed at 15\,Mbit/s and of 100 experiments with bandwidth ranging from 5 to 25\,Mbit/s (with bandwidth fixed at 15\,ms). All experiments were conducted for both the New Reno and Bic \glspl{cca}. This gives a total of 400 experiments, over which the values are averaged.} \label{tab:comparison_others}
\centering
\begin{tabular}{l r r r} \toprule
& \multirow{2}{*}{avg. throughp.} & \multicolumn{2}{c}{queue size} \\
& & max. & avg. \\ \midrule
\gls{ours}, offline $\alpha=0.01$ & 13.4 & 23.9 & 7.7\\
\gls{ours}, offline $\alpha=10$ & 12.5 & 12.7 & 3.4\\
\gls{ours}, online $\alpha=10$ & 12.8 & 16.1 & 4.5\\
FqCoDel	& 13.7 & 155.4 & 15.4\\
fq 100	& 11.7 & 100 & 51.1\\
fq 1000	& 11.9 & 1000 & 630.4 \\
\bottomrule
\end{tabular}
\end{table}

The systematic comparison (\autoref{tab:comparison_others}) between the different fair \gls{aqm} mechanisms shows that \gls{ours} achieves about the same or better throughput than the competing algorithms while having a significantly smaller average queue (less than half) and an even more pronounced reduction in maximum queue when compared to fq and FqCoDel.  

\section{Pitfalls of developing a \gls{dl}-based \gls{rl} systems for networks/telecommunications}

We believe that some implementation challenges that we encountered can be problematic also in other \gls{dl} based systems that use \gls{rl} to learn optimal behavior for networking and telecommunications. One thing that we found especially important is feature engineering: If features are fed as they are into the neural network, their quantity can be orders of magnitude apart. For example, if throughput is input as bits per second this is usually a number that is several millions or billions. On the other side, if queuing delay is measured in milliseconds, this means that it is a number somewhere between 1 and 1000 considering communications on Earth. This means that if inputs are not scaled to be of approximately the same size, neural network training fails either altogether (because of exploding gradients and numeric problems) or it is very unstable. Furthermore, smaller features are going to have a smaller influence on the output and it takes more training steps until they achieve the same importance as larger features as during each training step the neural network weights are just adjusted a little bit. This means that a small feature needs larger weights to achieve the same importance as a large feature and thus more training steps are required to grow the weights to the correct magnitude. For other \gls{dl} problems such as computer vision, this problem could be alleviated by normalization of the dataset. However, since in our scenario there is no fixed ``dataset'' but instead every network scenario that can happen in the real world has to be considered, normalization is not possible in our case because normalization always requires a pre-defined dataset. 

Another surprising insight was that simulating networks was significantly more resource-demanding than expected and took more computational resources than the \gls{dl}. Specifically, we simplified the network topology and even left out a switch connecting the hosts in our experiment topology as just having the switch would almost double the required simulation time for each flow. Thus, researchers should not overestimate the computation time required for \gls{dl} and not underestimate the one required for network/telecommunications simulation. This also implies that a GPU or other tensor processing device is not always required for \gls{dl} because it won't significantly speed up the training process. Furthermore, contrary to our expectation, that a network simulator can simulate flows faster than real time, this was not true: A flow with 15\,Mbit/s with a duration of 10\,s didn't take significantly less than 10\,s to simulate. Thus, network simulation is not suitable to speed up experiments compared to performing them on real hardware. This is especially true for very large bandwidth flows because the required simulation time increases with bandwidth because then more packets have to be processed. 

One major issue that we had during the implementation of \gls{ours} was that we encountered bufferbloat in our ns-3 simulation (over 100\,ms even if we set our buffers to 1 packet) and we couldn't pinpoint the location where it was introduced. The reason was that ns-3 has an additional buffer for each link: First, there's the buffer of the queuing discipline (for which we implement our mechanism) and then there's also a hardware buffer after that. This buffer has a default queue size of 100 packets. Thus, on a 10\,Mbit/s link, this hardware buffer alone results in a delay of 120\,ms when simulating bulk transfers. We reduced this buffer to 1 packet and then found the simulations to behave as expected. Thus, when using ns-3 for queuing experiments, it always has to be considered that by default there is a very large queue in each network interface, which not every user of ns-3 might be aware of. 

Last, the choice of the \gls{dl} toolkit is important. We started with tensorflow's C++ API but found it very difficult to make it work together with ns-3 due to tensorflow's unusual build system. On the other side, Pytorch's C++ API was easily linkable to ns-3 and also didn't require to be compiled from scratch, which saved compilation time. 

\section{Conlusion}

The results show that \gls{ours} learns the optimal buffering behavior taking into account the tradeoff parameter $\alpha$. This allows to tune how much throughput is allowed to be sacrificed for a reduction of average queue size. The online learning worked comparable to the offline learning, however, the online learning can take significantly longer to converge. This is no surprise as for offline learning we can do A/B testing between buffer sizing and thus always know what the optimal buffer size is. On the other side, for online learning, the learning depends on the output of a second neural network (the critic). This adds additional noise to the learning process. A way to alleviate this could be to pre-train a model using offline learning and then refine this pretrained model with online learning in a live deployment. 

Our comparison to existing \gls{aqm} methods shows that \gls{ours} achieves more or about the same throughput as other existing methods while at least halving the average queue size. The maximum queue size is decreased even more sharply: It is less than one fourth of the one of fq with a queue of 100 and less than a sixth of the maximum queue of FqCoDel. This observation is somewhat surprising as FqCoDel is specifically designed to prevent unnecessary long queues. However, while CoDel's algorithm manages to work reasonably well for flows in the congestion avoidance phase, it interacts badly with flows in the slow start phase. This is because during slow start, most \glspl{cca} increase the data in flight exponentially, which leads to a very quick increase in the queue. The problem is that CoDel uses an interval of 100\,ms to adjust the queue. Thus, slow start has 100\,ms to grow the queue exponentially and arbitrarily high. Only after that CoDel starts gradually decreasing the queue. This is particularly problematic as a large fraction of flows already end before slow start is over \cite{jurkiewicz_flow_2020} and thus could suffer from FqCoDel's large queue size. For example, for the average maximum queue size of 155 packets of FqCoDel shown in \autoref{tab:comparison_others} this means that on a 15\,Mbit/s link, the queuing delay alone is 124\,ms and can be even higher for smaller \glspl{rtt} as then the exponential increase of slow start can be even faster. 


We envision \gls{ours} being used on switches and routers close to the Internet's edge, as here usually the bottleneck links are and because network devices here only have to manage a rather small number of flows compared to the Internet's core. An interesting future direction of work could be to design and evaluate different reward functions, for example directly based on \gls{qoe}. 

\bibliographystyle{IEEEtranS}
\bibliography{reference}

\end{document}